\documentclass{aa} % for the letters 
\usepackage{amsmath}	% Advanced maths commands

\usepackage{graphicx}
\usepackage{xcolor}
\usepackage{txfonts}
\usepackage{natbib}
\usepackage{hyperref}
\hypersetup{
    colorlinks=true,
    citecolor=blue,
    linkcolor=blue,
    urlcolor=blue,}

\usepackage{longtable}
\usepackage{caption}

\makeatletter
\renewcommand*\aa@pageof{, page \thepage{} of \pageref*{LastPage}}
\makeatother

\bibpunct{(}{)}{;}{a}{}{,}

\newcommand{\bz}{\ensuremath{\langle B_z \rangle}}

\begin{document}

\title{Magnetically guided accretion and extremely slow rotation \\ in a  metal-enriched white dwarf}

\author{S.~Bagnulo\inst{1}
\and
M.~A.~Stroet\inst{1,2}
\and
C.~P.~Folsom\inst{3}
\and
M.~A.~Hollands\inst{4}
\and
J.~D.~Landstreet\inst{5,1}
\and
G. Ramsay\inst{1}
}

\institute{Armagh Observatory \& Planetarium, College Hill, Armagh BT61 9DG, UK
\and
School of Mathematics and Physics, Queen's University Belfast, University Rd, Belfast BT7 1NN, UK
\and
Tartu Observatory, University of Tartu, Observatooriumi 1, T\"oravere, 61602, Estonia
\and
Department of Physics, University of Warwick, Coventry CV4 7AL, UK
\and
Dept. of Physics \& Astronomy, University of Western Ontario, London, Ontario N6A 3K7, Canada
}

   \date{Received April 30, 2026; accepted July 22, 2026}

\abstract{
White dwarfs that show atmospheric metal enrichment offer a direct window into the bulk composition of exoplanetary material and provide some of the most direct constraints on the building blocks of rocky exoplanetary systems. Recent work has suggested that magnetic fields may influence how this material is accreted, potentially channelling debris along field lines and retaining it near magnetic poles rather than allowing it to be redistributed uniformly across the stellar surface. Here we report the discovery of photometric, spectroscopic, and spectropolarimetric variability in the magnetic DZ white dwarf WD\,1532+129, revealing a rotation period of about 289\,d, by far the longest yet directly measured for any degenerate star. The observed period may result from the combined effect of angular-momentum loss during the evolution leading to the white dwarf stage and an unusually slowly rotating progenitor, perhaps a magnetic Ap star. The white dwarf exhibits surface abundance inhomogeneities, with metals concentrated near both magnetic poles, and the observed modulation is consistent with localised metal-rich regions aligned with the magnetic geometry. WD\,1532$+$129 is the third of the four known magnetic DZ white dwarfs within 20\,pc to show evidence for polar metal enhancement, indicating that magnetically guided inhomogeneities are a common property of magnetic metal-enriched white dwarfs.
}

\maketitle

%
%-------------------------------------------------------------------

\section{Introduction}
White dwarfs frequently bear the hallmarks of their long-gone planetary systems, including photospheric traces of heavy elements accreted from circumstellar debris disks \citep{Zucetal03,Koeetal14}. These metal-enriched white dwarfs provide a direct window into the bulk composition of exoplanetary material, offering constraints on the building blocks of rocky planetary systems and their dynamical evolution \citep{Zucetal03,Faretal09}. In most cool white dwarfs, efficient surface convection is expected to homogenise freshly accreted metals over the stellar photosphere, producing a chemically uniform surface distribution on timescales much shorter than those for accretion and gravitational settling \citep{Cunetal21}. However, recent observations challenge this expectation.

Magnetic fields are known to play a central role in stellar surface chemistry and structure, and may similarly govern how accreted planetary material is distributed across the surface of white dwarfs. In particular, a weak magnetic field can truncate a gaseous accretion disk and guide infalling plasma along field lines toward the magnetic poles \citep{Metetal12,Faretal18}. The first clear example of such magnetically channelled accretion was identified in the nearby DZ white dwarf WD\,0816–310, whose $\sim 140$\,kG field confines metals to a persistent “scar” near a magnetic pole, producing synchronous variations in both the magnetic field and metal line strength as the star rotates \citep{Bagetal24a}. A second object, WD\,2138$-$332, shows the same behaviour: a clearly variable $\sim 60$\,kG dipolar field, periodic changes in metal line strength, and correlated variation of the longitudinal field, demonstrating that even a modest field can shape the surface pattern of accreted material \citep{Bagetal24b}. These objects are two of the four magnetic DZ white dwarfs in the 20 pc volume around the Sun.

The discovery of these two systems established a new physical picture for metal accretion onto white dwarfs, in which even weak magnetism can channel ionised debris streams onto small surface regions. Despite the recognition that magnetic fields can shape the accretion flow of planetary debris, the overall frequency and physical consequences of magnetically channelled accretion remain poorly constrained. Surface inhomogeneities can only be revealed through rotational modulation, providing a probe of both the magnetic geometry and the distribution of accreted material. However, only a few metal-enriched white dwarfs with measurable magnetic fields have been monitored in sufficient detail to test for abundance inhomogeneities, and even fewer have been observed with phase-resolved spectropolarimetry. Rotational modulation also provides one of the few direct ways to measure the spin periods of these stars, and thus to test whether magnetic, metal-enriched white dwarfs show unusual rotational behaviour. Establishing whether the surface distribution of metals is commonly linked to magnetic geometry, and determining how magnetism alters the transport and retention of heavy elements within the atmosphere, are essential next steps. In this work, we present new spectropolarimetric observations of a third cool, weakly magnetic DZ white dwarf within 20 pc, combined with archival photometry. Our analysis allows us to investigate both the surface distribution of accreted metals and the rotational behaviour of this system.

\section{Observations}
%%%%%%%%%%%%%%%%%%%%%%%%%%%%%%%%%%%%%%%%%%%%%%%%%%%%%%%%%%%%%%%%%%%%%%%%%%%%%%%%%%%%%%
\begin{table*}
\centering
\caption{
Observing log and longitudinal magnetic field measurements of WD\,1532+129.
}
\label{Table_Log}
\begin{tabular}{lllccr@{\,$\times$\,}lrr@{\,$\pm$\,}lc}
\hline\hline
INSTR. & Grism/        & MJD    & DATE       & UT    & \multicolumn{2}{c}{Exp} & S/N        & \multicolumn{2}{c}{\bz} & REF. \\
       & Grating       &        & yyyy-mm-dd & hh:mm & \multicolumn{2}{c}{(s)} & \AA$^{-1}$ & \multicolumn{2}{c}{(kG)} &      \\
\hline
FORS2 & 1200B & 58560.271 & 2019-03-18 & 06:30 & 4 & 675 & 230 & $-5.0$  & 1.4 & \citet{BagLan19b} \\
ISIS  & R600B & 58592.106 & 2019-04-19 & 02:33 & 8 & 600 &  55 & $-18.8$ & 7.3 & \citet{BagLan19b} \\
FORS2 & 1200B & 58653.101 & 2019-06-19 & 02:26 & 4 & 660 & 225 & $-23.1$ & 2.1 & \citet{BagLan19b} \\
FORS2 & 1200B & 60768.353 & 2025-04-03 & 08:28 & 4 & 740 & 285 & $+24.1$ & 1.0 & this work \\
FORS2 & 1200B & 60770.347 & 2025-04-05 & 08:19 & 4 & 740 & 295 & $+24.4$ & 0.9 & this work \\
FORS2 & 1200B & 60788.301 & 2025-04-23 & 07:13 & 4 & 740 & 250 & $+26.4$ & 1.2 & this work \\
FORS2 & 1200B & 60792.324 & 2025-04-27 & 07:47 & 4 & 650 & 280 & $+25.6$ & 0.9 & this work \\
FORS2 & 1200B & 61148.183 & 2026-04-18 & 04:23 & 4 & 676 & 270 & $+1.6$  & 1.1 & this work \\
\hline
\end{tabular}
\tablefoot{ The \bz{} estimates were obtained from the wavelength ranges 4196--4254\,\AA\ and 4359--4405\,\AA,
adopting $g_{\rm eff} = 1.25$.}
\end{table*}
%%%%%%%%%%%%%%%%%%%%%%%%%%%%%%%%%%%%%%%%%%%%%%%%%%%%%%%%%%%%%%%%%%%%%%%%%%%%%%%%%%%%%%

\subsection{Spectropolarimetry}
A spectropolarimetric survey of nearby white dwarfs found that four of the 13 DZ/DZA white dwarfs within the local 20\,pc volume have a magnetic field: WD\,0816$-$310, WD\,1009$-$184, WD\,1532$+$129 and WD\,2138$-$332 \citep{BagLan19b}. After the discovery of absorption line strength variability due to the presence of metal abundance patches at the surface of the DZ stars WD\,0816$-$310 \citep{Bagetal24a} and WD\,2138$-$332 \citep{Bagetal24b}, we monitored a third known magnetic DZ white dwarf of the local 20\,pc volume, WD\,1532+129. Its effective temperature, mass, and cooling age are 5796\,K, $0.63\,M_\odot$, and 4.0\,Gyr, respectively \citep{OBretal24}.

Observations were obtained with the FORS2 instrument \citep{Appetal98} of the ESO VLT, using the 1200B grism which covers the wavelength range 3700--5200\,\AA. We used a 1\arcsec\ slit width, for a resolving power of 1400. We obtained new observations at five different epochs in 2025 and 2026. Each observing series consisted of two pairs of exposures. The same star had been observed twice in 2019 with FORS2 using the 1200B grism (again with a 1\arcsec\ slit width), and once in 2019 with ISIS, using the R600B grating \citep{BagLan19b}.

Data reduction was carried out with the FORS2 pipeline \citep{Izzetal10} to obtain 2-dimensional wavelength-calibrated frames, from which beams were extracted using standard procedures within {\sc IRAF}. Beam recombination was performed with simple {\sc Fortran} routines. The intensity spectra were divided by the intensity spectrum of the white dwarf EC 20173-3036, a featureless He-rich white dwarf that was observed with the same setting both on 2019-06-19 and on 2025-10-29. Before being used for the normalisation, the two Stokes~$I$ spectra of EC~20173-3036 were smoothed with a 10-pixel boxcar. The mean longitudinal field \bz\ was measured at all epochs with the technique used by \citet{BagLan19b}. As discussed by \citet{Lanetal14}, \bz\ estimates depend on the instrument used and on the wavelength interval adopted for the measurement, therefore the uncertainties of our measurements are meaningful only in the context of internal consistency. \bz\ was determined from the ranges 4196 to 4254\,\AA\ and 4359 to 4405\,\AA. In all cases we adopted an effective Land\'e factor $g_{\rm eff} = 1.25$. The observing log is given in Table~\ref{Table_Log} and includes the measurements previously published, which were re-reduced to adopt exactly the same wavelength intervals specified above. 

We also calculated equivalent widths (EWs) spanning multiple (blended) spectral features.  We used the wavelength ranges $4357$--$4421$\,\AA\ and $4180$--$4335$\,\AA, and also a wider window spanning $4162$--$4466$\,\AA\ (we avoided the range below 4100\,\AA, because the continuum level is quite uncertain). Results are given in Table~\ref{Table_EW_WD1532} of App.~\ref{App_Log}.

\subsection{Photometry}
It is known that some magnetic WDs exhibit photometric variability from which the rotational period may be inferred. WD\,1532+129 has been checked for variability with periods up to several days using TESS data, but no variation was detected \citep{Heretal24}. We therefore explored the possibility that photometric surveys with much slower cadence than TESS might reveal longer period variability, and retrieved data from three imaging photometric surveys. GOTO has sixteen 40 cm telescopes in both La Palma, Canaries, and Siding Spring, Australia and typically uses an $L$ filter (400--700 nm) \citep{Lyman2026}; ATLAS has two 0.5 m telescopes in Hawaii, with identical 0.5 m telescopes in Chile and South Africa \citep{Tonry2018,Heinze2018} with observations made in the `c' filter (420--650 nm) and `o' filter (560--820 nm); and ZTF has a 1.2\,m and a 1.5\,m telescope in Palomar, California, and uses $g$ (409--552 nm), $r$ (560--732 nm) and $i$ (688--901 nm) filters \citep[for a full description see][]{Bellm2019}. All three surveys have cadences of the order of one or a few days, and so are sensitive to variations with time scales of a few days or longer. 
The earliest photometric data on WD\,1532+129 were obtained in ATLAS from August 2015, ZTF from March 2018, and in GOTO from Sept 2023. WD\,1532+129 has a significant proper motion \citep{GaiaDR3} that had to be taken into account when extracting the forced photometry. ATLAS data were filtered to remove bad data points using the filtering recommended on the ATLAS website. 

\section{Analysis of the stellar variability}
%%%%%%%%%%%%%%%%%%%%%%%%%%%%%%%%%%%%%%%%%%%%%%%%%%%%%%%%%%%%%%%%%%%%%%%%%%%%%%%%%%%%%%%%%%%%%
\begin{figure*}
\begin{center}
\includegraphics[angle=0,width=8.8cm,trim={0.8cm 5.5cm 1.1cm 3.0cm},clip]{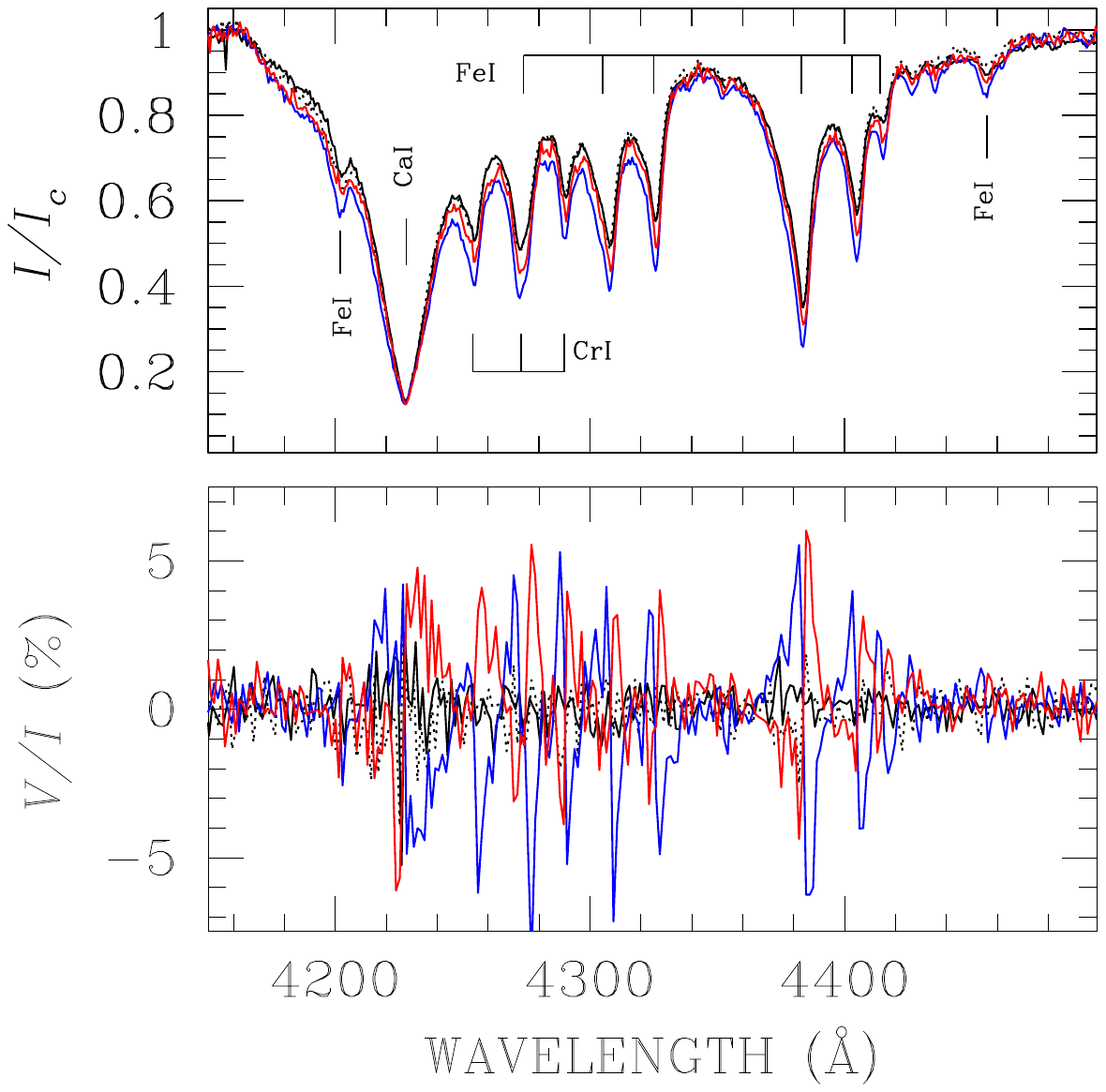}
\includegraphics[angle=0,width=8.8cm,trim={0.8cm 5.5cm 1.1cm 3.0cm},clip]{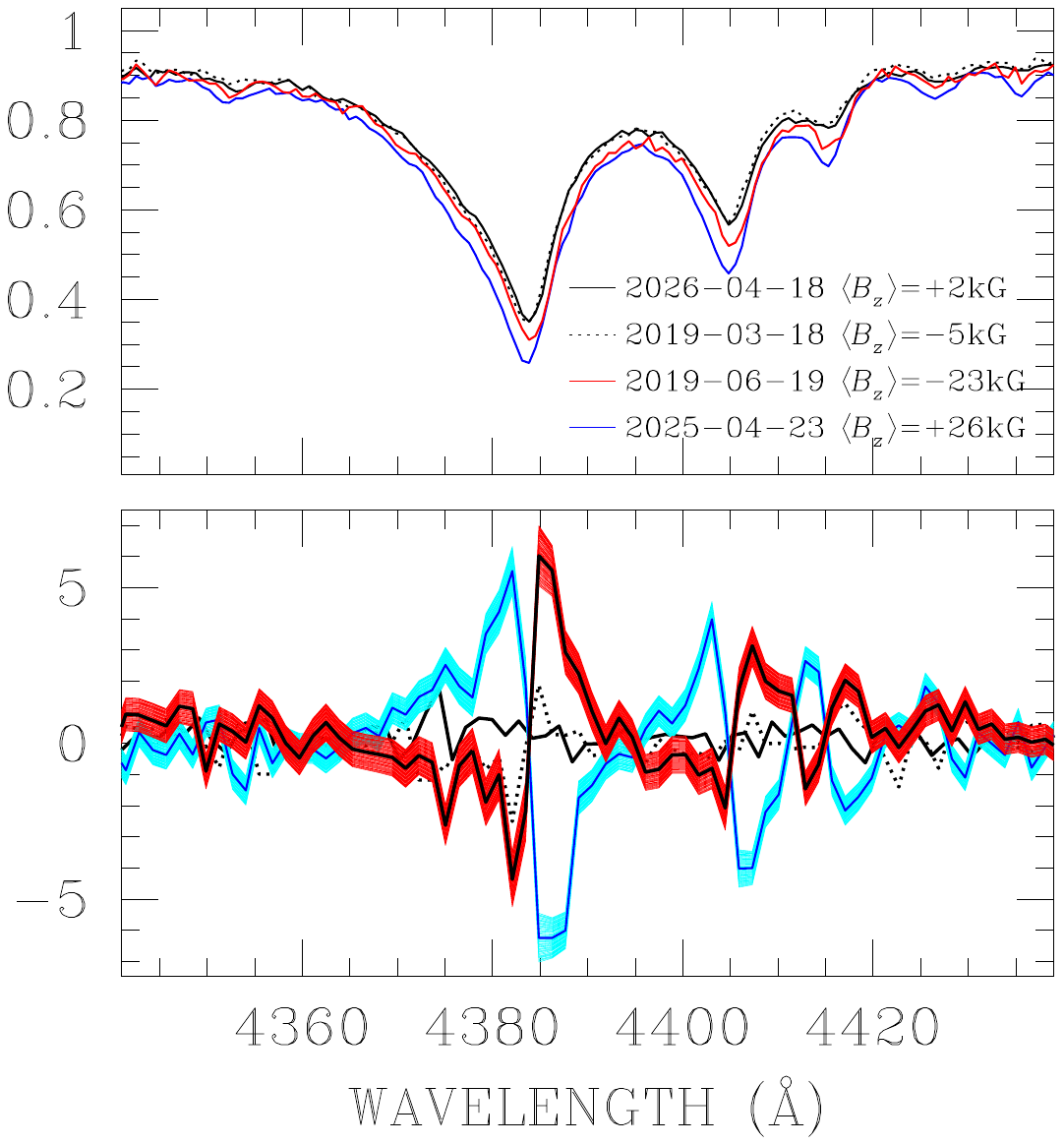}
\end{center}
\caption{\label{Fig_WD1532} Intensity $I$ normalised by the continuum $I_{\rm c}$, and reduced polarised spectra (Stokes $V/I$) of WD\,1532$+$129 obtained at four different epochs. The two right panels show an enlarged view of the two left panels. The red lines correspond to a time when the mean longitudinal magnetic field has a positive maximum ($\sim 26$\,kG), and the star is viewed approximately magnetic-pole-on; the black lines correspond to epochs when the mean longitudinal field is nearly zero and the star is viewed approximately magnetic-equator-on; the blue lines correspond to an epoch when the negative magnetic pole is in the visible stellar hemisphere as the mean longitudinal field is negative ($\sim -23$\,kG). The shaded regions correspond to $1\,\sigma$ uncertainties. Intensity spectra are displayed with the original $\sim 0.7$\,\AA\ bin, while Stokes $V/I$, to reduce the noise, are rebinned at $\sim 1.4$\,\AA.
}
\end{figure*}
%%%%%%%%%%%%%%%%%%%%%%%%%%%%%%%%%%%%%%%%%%%%%%%%%%%%%%%%%%%%%%%%%%%%%%%%%%%%%%%%%%%%%%%%%%%%%

Multi-epoch intensity and circular polarisation spectra of WD\,1532$+$129 obtained with FORS2 are shown in Fig.~\ref{Fig_WD1532}. The strong Ca\,{\sc i} line shows variable circular polarisation profiles, and nearly constant intensity profiles, and so do the even deeper Ca\,{\sc ii}\,H\&K lines (not shown in the figure). Weaker Fe\,{\sc i} lines show variability both in intensity and in circular polarisation. These line strength changes are more subtle than those observed in WD\,0816$-$310 and in WD\,2138$-$332 \citep{Bagetal24a,Bagetal24b}. However, the observed variability exceeds the noise.

Magnetic fields are known to increase the equivalent widths of spectral lines, thereby mimicking enhanced chemical abundance \citep{Babcock49,StiLeo03}. We performed numerical experiments with the Zeeman code \citep{Landstreet88,Lanetal17}, using the same He-rich atmosphere model adopted by \citet{Holetal22} for their spectral analysis of WD\,1532+129. We considered simple magnetic configurations with surface field strengths ranging from 0 to 300\,kG. We found that the observed variations in line shape and equivalent width remain far larger than can be explained by rotationally modulated changes in magnetic intensification. This conclusion holds even when comparing a zero-field model with a model field strength of 300\,kG; in the latter case, magnetic splitting would already begin to be detectable at the spectral resolution of FORS2. In App.~\ref{App_Scattered} we also consider and rule out the hypothesis that the observed changes in the line equivalent width may be an artefact due to scattered light, and we show through radiative transfer calculations that metal abundance changes of the order of 0.2\,dex can explain the observed variability of the line strength of the weaker lines, while keeping the profile of the strongest lines nearly constant. In the following, the line strength variability will be interpreted in terms of a non-homogeneous distribution of the metals at the surface of the star.

Our measurements reveal that the longitudinal field of WD\,1532$+$129 reverses sign (see Table~\ref{Table_Log}), but the general consistency between the four spectra obtained from April 3 to 27 2025 suggests that the stellar rotation period may be either roughly a multiple of 1 day, or at least several months long. 

Remarkably, period analyses of the ZTF, ATLAS and GOTO photometry all show periodic variability with a period of approximately 144\,d. To evaluate whether this photometric variability was due to systematic trends, we extracted ATLAS photometry for three nearby comparison stars with $G\sim16$ within 2\arcmin\ of WD\,1532+129, applying the same filtering criteria to their light curves. We found no evidence of significant periodic signals in any of the three comparison stars. 

WD\,1532+129 is a single WD according to \citet{Tooetal17}, which suggests that periodic light variability is likely linked to stellar rotation. However, the 144\,d period was found inconsistent with the observations of the mean longitudinal field (see Fig.~\ref{Fig_WD1532_all144d} in App.~\ref{App_Period}). Since WD\,1532$+$129 reverses the sign of its observed longitudinal field, the two magnetic hemispheres must be seen at different rotational phases. In such a geometry, the photometric modulation may repeat twice per rotation, so that the dominant photometric period corresponds to one half of the true rotation period $P$, as observed in other magnetic white dwarfs such as WD\,0009+501 and WD\,0041--102 \citep{Heretal24}. In fact, we found that adopting twice the dominant photometric period as the rotation period and fitting each of the six photometric light curves and three EW curves with a second-order harmonic model,
\begin{equation*}
    f_i(t) = A_i + B_i\sin(\omega t) + C_i\cos(\omega t) + D_i\sin(2\omega t) + E_i\cos(2\omega t)\; ,
\end{equation*}
where $\omega=2\pi/P$, and $i$ labels the individual data series, while fitting the longitudinal field measurements with
\begin{equation*}
    {\left<B_z\right>}\,(t) = A' + B'\sin(\omega t) + C'\cos(\omega t),
\end{equation*}
yielded a fully consistent fit with a best-fitting period $P=288.7 \pm 0.9$ days.

\begin{figure}[!h]
\begin{center}
\includegraphics[width=\columnwidth,trim={0.25cm 0.20cm 0.25cm 0.1cm},clip]{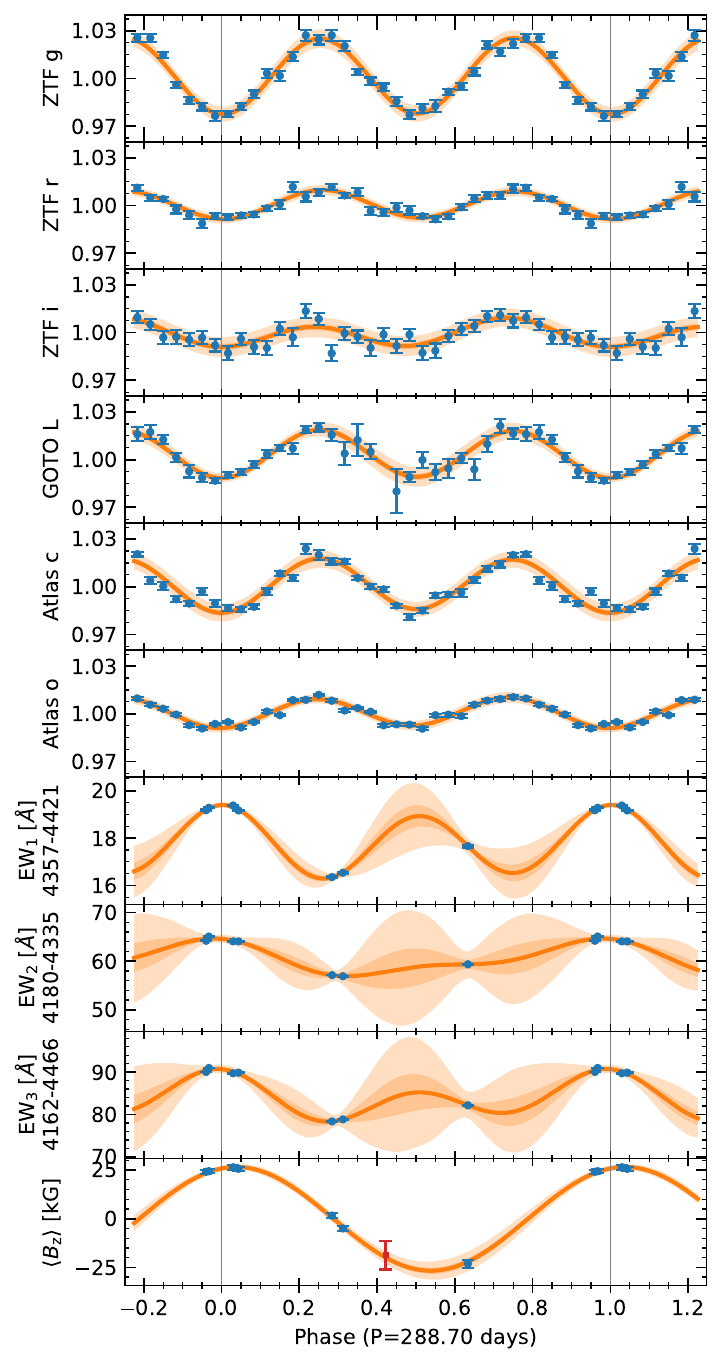}
\end{center}
\caption{\label{Fig_WD1532_All} From top to bottom: WD\,1532+129 photometry from ZTF, GOTO, and ATLAS; equivalent width measurements from FORS2 spectra; and mean longitudinal field values, all phased with a rotational period of 288.7\,d and a zero-point of MJD 59047.566. Solid curves show the best-fit models, while orange shaded regions mark the 1\,$\sigma$ and 3\,$\sigma$ confidence intervals. In the bottom panel, the red point with a visible error bar is the \bz\ estimate obtained with ISIS at the WHT. It was not included in the fit, but is fully consistent with the long period derived here. The error bars of the FORS2 equivalent width and \bz\ measurements are smaller than the symbols.  The vertical lines delimit one complete rotational cycle.}
\end{figure}

All WD\,1532+129 light curves, EW curves, and the longitudinal-field curve are shown in Fig.~\ref{Fig_WD1532_All}, phased with a rotational period of 288.7\,d. The adopted zero-point, MJD 59047.566, corresponds to the minimum of the ZTF g-band light curve. The orange curves show the corresponding best-fitting models, while the shaded regions represent the 1\,$\sigma$ and 3\,$\sigma$ confidence intervals derived from 10\,000 Monte Carlo realisations of the best-fitting parameters.

We note that the use of a sinusoidal model for $\langle B_z\rangle$ is not an assumption that the field of WD\,1532+129 has a simple morphology. The mean longitudinal field is a disk-averaged observable and is therefore most sensitive to the lowest-order, large-scale components of the magnetic field. More complex components tend to cancel in the disk average. For example, the contribution of a linear quadrupolar component to $\langle B_z\rangle$ is much smaller than that of a comparable dipolar component \citep{Schwarz50}, and this conclusion is also valid for a planar quadrupole \citep{Bagetal96}. Thus, even a complex field will produce an approximately sinusoidal longitudinal-field curve dominated by the dipolar component. Strongly non-sinusoidal behaviour would require a higher-order component large enough to survive disk averaging, in which case it would likely also leave detectable signatures in Stokes~$I$, through magnetically split line profiles.

%%%%%%%%%%%%%%%%%%%%%%%%%%%%%%%%%%%%%%%%%%%%%%%%%%%%%%%%%%%%%%%%%%%%%%%%%%%%%%%%%%%%%%%%%%%%
%%%%%%%%%%%%%%%%%%%%%%%%%%%%%%%%%%%%%%%%%%%%%%%%%%%%%%%%%%%%%%%%%%%%%%%%%%%%%%%%%%%%%%%%%%%%

\section{Magnetic field and metal abundance patches}\label{Sect_Discussion}
To investigate the relationship between magnetic variability and line-strength variability, it is useful first to infer an approximate picture of the magnetic field geometry from the observations. 

The longitudinal magnetic field of WD\,1532+129 varies periodically and reverses sign as the star rotates. Since the longitudinal field is primarily sensitive to the dipolar component of the magnetic field, the observed sign reversal implies that both magnetic poles of the dipole become visible during a rotation cycle. 
The fact that the phased variation of $\bz$ is nearly symmetric about $\bz = 0$ indicates either that the rotation axis is approximately perpendicular to the line of sight (while the dipolar axis may be inclined by any angle $\beta$ with respect to the rotation axis), or that the dipolar axis is nearly perpendicular to the rotation axis (while the rotation axis may be inclined by any angle $i$ with respect to the line of sight).

Because the maximum observed value of $\vert\bz\vert$ is approximately 25\,kG, the polar strength $B_{\rm p}$ of a dipolar field must be at least $\sim 80$\,kG. The absence of detectable Zeeman splitting in the low-resolution spectra implies that the mean field modulus is $\la 300$\,kG, which in turn formally sets an upper limit of approximately 450\,kG on $B_{\rm p}$. Assuming a centred dipole and adopting, somewhat arbitrarily, $i = 90^\circ$ and $\beta = 60^\circ$, one obtains $B_{\rm p} \sim 100$\,kG \citep[see, e.g., Eq.~1 of][adopting a limb darkening coefficient $u=0.5$]{Preston67}. For $i = 90^\circ$ and $\beta = 30^\circ$, the corresponding polar strength is about 160\,kG (we recall that the expression for $\bz$ is symmetric with respect to the interchange of $i$ and $\beta$). Overall, these considerations suggest that it is plausible to assume dipolar field strength at the magnetic poles of order 100--150\,kG. Stronger constraints could be obtained from high-resolution spectroscopy of the Fe\,{\sc i} lines.

The EWs of the Fe~{\sc i} lines vary with magnetic phase: the lines become deeper when either magnetic pole is close to the line of sight, indicating that the local abundance of Fe is enhanced in the polar regions relative to the magnetic equator, confirming that metals are more concentrated where the magnetic field lines are locally perpendicular to the stellar surface.

WD\,1532+129 is the third DZ white dwarf in which metal abundance patches have been found to be associated with the magnetic poles. The same phenomenon was previously discovered in two other magnetic DZ white dwarfs within the local 20\,pc volume: WD\,0816--310 \citep{Bagetal24a} and WD\,2138--332 \citep{Bagetal24b}. 

The available photometric bands cover most of the flux emitted by WD\,1532+129, from $\sim 400$\,nm to beyond 900\,nm. The light curves are approximately anti-correlated with the Fe\,{\sc i} equivalent widths, so that the star is faintest near magnetic pole-on phases. All available light curves vary approximately in phase, implying that the variability is not caused by simple wavelength redistribution from blue to red and infrared bands. Instead, the total local surface flux appears to be lower near the magnetic poles than at the equator. The same behaviour was observed in WD\,2138$-$332 \citep{Bagetal24b}, and remains at odds with the expectation that the local surface flux of a magnetic white dwarf should be approximately constant over the stellar surface. Although magnetic fields can suppress convection and thereby modify the atmospheric structure, theoretical models show that suppressing convection does not significantly affect the global cooling rate, at least for white dwarfs with $T_{\rm eff} \gtrsim 5500$\,K \citep{Treetal15}. 

No photometric variability has so far been detected in WD\,0816--310 \citep{Faretal24}. 

The three DZ stars WD\,0816--310, WD\,1532+129, and WD\,2138--332 may be regarded as cooler, metal-polluted analogues of a number of DB/DAB white dwarfs that exhibit both magnetic variability and surface-composition inhomogeneities, such as, for instance, Feige\,7, PG\,0853+164, the ``two-faced'' white dwarf ZTF\,J203349.8+322901.1 (nicknamed Janus), and SDSS\,J091016.43+210554.2 \citep{Achetal92,Wesetal01,Caietal23,Mosetal24}. The recent work of \citet{Mosetal25} confirms that these and other similar objects form part of an emerging class of double-faced white dwarfs, in which the observed H/He line-strength variations are interpreted in terms of non-uniform surface composition, often associated with magnetic-field geometry.

For Janus, \citet{Caietal23} discussed several possible magnetic explanations for the observed H/He surface inhomogeneity. First, the magnetic field may locally suppress convection, allowing one region to preserve a hydrogen-rich layer while another is mixed and becomes helium-dominated. Secondly, the magnetic field may produce small differences in temperature or pressure structure across the surface; because the onset of helium convection is very sensitive to the local atmospheric conditions, such differences may be sufficient to produce different apparent H/He compositions on different parts of the star. Thirdly, magnetic pressure may contribute directly to horizontal chemical separation in a stratified atmosphere. Similarly, \citet{Mosetal25} modelled several double-faced white dwarfs with hydrogen-rich polar caps and helium-rich equatorial regions, and argued that magnetism is likely to affect the convective processes that would otherwise lead to a more homogeneous H/He distribution. 

These mechanisms for producing surface inhomogeneities generally rely on the fact that hydrogen, as a trace element, floats to the stellar surface. The situation in cool DZ white dwarfs is different. In these stars, the inhomogeneous species are not residual elements involved in spectral evolution, but metals accreted from tidally disrupted planetary material \citep{Jura03,Faretal09}. In the absence of a magnetic field, these metals are expected to be mixed vertically throughout the convection zone, redistributed horizontally over the stellar surface, and eventually depleted by gravitational settling \citep[e.g.][]{Koester09,Koeetal14}. For cool helium-rich atmospheres, theoretical calculations predict efficient horizontal homogenisation on a vertical diffusion timescale \citep{Cunetal21}. The same calculations also suggest that a field of order 100\,kG is not, by itself, sufficient to suppress convection completely in cool white-dwarf atmospheres \citep{Cunetal21}. This makes the persistence of metal-line strength patches in WD\,1532+129, WD\,0816--310, and WD\,2138--332, all of which have field strengths of order 100--200\,kG, particularly significant.

Since the close correlation between magnetic-field geometry and metal-line strength strongly suggests a causal role for the magnetic field, \citet{Bagetal24a} proposed that the magnetic field may channel ionised accretion flows preferentially towards the magnetic poles, producing an intrinsically non-uniform deposition pattern. The field may also inhibit, or at least make anisotropic, the subsequent horizontal redistribution of accreted material. While it is known that weak fields can truncate an ionised gas disk and guide the accretion stream towards the surface near each magnetic pole \citep{Metetal12}, a difficulty with the channelling scenario is that cool DZ white dwarfs do not emit enough UV flux to ionise the accreting gas. However, \citet{Bagetal24a} argued that close to the white dwarf even a tiny initial population of ions would be forced to circle around the magnetic field lines. These ions would move at high velocity relative to the neutral gas and could ionise additional atoms through collisions, triggering a rapid ionisation cascade. In this picture, once the inflowing gas becomes sufficiently ionised, it is naturally guided along the magnetic field lines and deposited preferentially near the magnetic poles. While this provides a plausible mechanism for producing metal-rich polar regions, the relative importance of magnetically channelled accretion, inhibited horizontal mixing, and local atmospheric-structure effects remains uncertain.

\section{Slow rotation}
Using TESS data, \citet{Olietal24} found median rotation periods of 3.4\,h for confirmed magnetic white dwarfs and 1.9\,h for the subset of single systems. Using both photometric and polarimetric data, \citet{Heretal24} reached a similar conclusion, while also drawing attention to a few objects with periods of several days. Both samples include, and probably over-represent, ultra-massive white dwarfs, which are likely fast-rotating merger products and may follow a different rotational evolution from single-star remnants such as WD\,1532+129. Restricting the comparison to average-mass white dwarfs, for instance those with $M \le 0.75\,M_\odot$, still yields predominantly short periods, with median values of about 4--5\,h. Non-magnetic white dwarfs appear to rotate more slowly than magnetic white dwarfs \citep{Heretal24,Olietal24}, but their typical periods \citep[25--35\,h,][]{Heretal17} are still no longer than 1 or 2\,d. WD\,1532+129 therefore remains an extreme outlier, with by far the longest rotation period measured for any white dwarf, magnetic or not: $P \simeq 289$\,d is about 16 times longer than the previous record holder, $P = 17.86$\,d of WD\,2316+123 \citep{SchNor91}.

A class of strongly magnetic, polarimetrically non-variable white dwarfs has been identified in the literature and is often interpreted as consisting of extremely slowly rotating stars, with field strengths of tens of MG and rotational periods of the order of centuries \citep[e.g.][]{Kawka20,Feretal20}. \citet{BagLan24} gave a different interpretation, noting that no white dwarfs have been observed with rotational periods longer than about 18\,d, and arguing that it is therefore unlikely that magnetic white dwarfs simply spin down through this interval. They instead proposed that the absence of polarimetric variability in these objects is best explained by the alignment of the magnetic and rotational axes in old, strongly magnetic white dwarfs. In this context, WD\,1532+129 might appear to provide a direct observational bridge between rapidly rotating and extremely slowly rotating white dwarfs, potentially challenging the conclusion of \citet{BagLan24}. However, WD\,1532+129 is quite old \citep[$\sim 4.0$\,Gyr,][]{OBretal24} and only weakly magnetic ($\sim 0.1$\,MG), and therefore does not fit naturally into a scenario in which strongly magnetic white dwarfs undergo substantial spin-down. Its extreme rotation period may therefore have a different origin, rather than representing a transitional stage between fast and ultra-slow rotators.

Detecting very long rotational periods photometrically is intrinsically difficult. Many time-series data sets, including those obtained by missions such as TESS, are detrended to remove instrumental systematics and long-term drifts, a process that can attenuate or suppress genuine long-period signals. The TESS-based sample of \citet{Olietal24} is naturally biased towards fast rotators, and one might speculate that a population of slowly rotating white dwarfs remains undetected. However, this effect alone is unlikely to explain the apparent scarcity of such objects. Polarimetric observations, which often span several years, are sensitive to much longer timescales and should reveal rotational modulation unless that modulation is intrinsically suppressed. If one considers only the average-mass stars in Table~C.1 of \citet{Heretal24} whose rotation periods were not inferred via photometry but from magnetic measurements, the distribution still favours short periods, with a median of only 2.4\,h. The mean period is longer, about 28\,h, but it is strongly influenced by a small number of longer-period objects. In conclusion, the available evidence indicates that magnetic white dwarfs are generally rapid rotators, with typical periods of the order of minutes to hours, but that objects with periods up to several days are not uncommon. However, objects as slow as WD\,1532+129 are unlikely to constitute a large hidden population.

We note that WD\,0816$-$310, the first white dwarf in which magnetically guided accretion was identified, is a relatively slow rotator, with a period longer than the typical few-hour range. No light variability has been detected in this star, and the existing \bz\ and EW measurements do not tightly constrain its period. Nevertheless, they are consistent with values between about 2 days and 2 weeks \citep{Bagetal24a}.  Magnetic white dwarfs with similarly long rotation periods are not uncommon. This could prompt speculation that slower rotation may be driven by interactions between the stellar magnetic field and circumstellar debris, either at present or during earlier episodes of planetary accretion. Such a mechanism would be analogous to disk locking in T Tauri stars \citep{Koenigl91,Lonetal05,Bouetal07} or to the spin evolution of magnetic accreting neutron stars \citep{GhoLam79}. However, accretion rates in metal-enriched white dwarfs are many orders of magnitude lower than in these systems. Moreover, for WD\,1532$+$129, the corotation radius,
\[
r_{\rm co}=\left(\frac{GMP^2}{4\pi^2}\right)^{1/3},
\]
corresponding to a rotation period of \(\sim 289\) d would lie at \(\sim 0.7\) au, more than \(10^4\) stellar radii from the white dwarf. At that distance, a dipolar surface field of order \(0.1\) MG would be only a few times $10^{-8}$\,G, rendering efficient magnetic locking implausible. 
For a representative polluted-white-dwarf accretion rate we adopt the value of \(\dot M = 10^8\) g\,s\(^{-1}\), chosen conservatively to be smaller than the lowest value estimated by \citet{Swaetal23} for six objects. For such a value, the spherical Alfv\'en radius
\[
r_{\rm A}\simeq \left(\frac{\mu^4}{2GM\dot M^2}\right)^{1/7},
\]
with \(\mu \simeq B_{\rm eq}\, R_*^3 = B_{\rm p}\, R_*^3/2\), would be only of order 0.06\,au ($\sim 1100$ stellar radii); % \(10^{12}\)\,cm; 
the actual disk truncation radius would be even smaller, and certainly far inside corotation. In that regime the interaction would more naturally spin the star up rather than brake it. Furthermore, the above estimate assumes efficient coupling between the stellar magnetic field and the circumstellar material, while debris disks around cool white dwarfs can contain a substantial neutral or dusty component rather than a fully ionised accretion flow. Any reduction in magnetic coupling would further weaken the case for disk locking. Even if star--disk coupling were stronger at earlier, more rapidly rotating stages, the low accretion rates and limited debris masses of polluted white dwarfs make it extremely unlikely that this mechanism could remove enough angular momentum to produce a period of order one year.

The rotation of white dwarfs is generally understood to reflect substantial angular-momentum redistribution and loss during post-main-sequence evolution. Estimates based on the change in stellar structure between the main sequence and the white-dwarf stage predict remnant spins much faster than those actually observed \citep{Spruit98,Koeetal98,Heretal17}. This has long pointed to the need for efficient angular-momentum transport and loss during later evolutionary phases. In this context, WD\,1532$+$129 appears to represent an especially extreme outcome of this general evolutionary problem. A plausible interpretation is that WD\,1532$+$129 is the descendant of a slowly rotating progenitor, for instance a magnetic Ap star. Ap stars are known to rotate significantly more slowly than normal A-type stars, with periods of months not uncommon \citep{Netetal17}, and a very small subset reaches periods of decades or longer; for instance, the roAp star $\gamma$\,Equ likely rotates with a period of about 97\,yr \citep{Bycetal16}, while HD\,50169 has a securely established period of 29\,yr \citep{Matetal19}. Thus, the problem of very large angular-momentum loss is already present on the main sequence. Even a progenitor of this kind, however, would still need to lose most of its angular momentum before reaching the white-dwarf stage. The exceptionally long rotation period of WD\,1532$+$129 may therefore reflect the combined effect of unusually slow progenitor rotation and substantial angular-momentum loss during later evolution.

\section{Conclusions}
Within the local 20\,pc volume, there are 13 known DZ/DZA white dwarfs, and four of them have a weak but firmly detected magnetic field. 
At least three of the four known magnetic DZ white dwarfs within the local 20\,pc volume exhibit spectral line variability synchronised with the rotational modulation of the mean longitudinal field, indicating that the photospheric abundances of heavy elements are enhanced near the magnetic poles and reduced at lower magnetic latitudes. This group comprises WD\,0816$-$310 \citep{Bagetal24a}, WD\,2138$-$332 \citep{Bagetal24b}, and the newly identified here as a third such case WD\,1532$+$129, which is the first one in which metal enhancement can be observed at both magnetic poles. These results indicate that magnetically guided accretion is common among magnetic, metal-enriched white dwarfs. 

Remarkably, WD\,1532$+$129 has by far the longest rotational period ever measured in a white dwarf, $P \sim 289$\,d, an order of magnitude longer than the longest rotational period measured before. This extreme rotation period is unlikely to result from magnetic spin-down during the cooling phase. Slow rotation of white dwarfs in general already implies substantial angular-momentum loss during earlier evolution, but the exceptional period of WD\,1532+129 may represent an extreme case in which this general evolutionary effect is further enhanced by descent from an unusually slowly rotating progenitor, perhaps a magnetic Ap star.

Beyond the specific case of the exceptionally slow rotation of WD\,1532$+$129, our results highlight the importance of understanding how magnetic fields influence the retention and transport of metals in white dwarf atmospheres. The time that heavy elements persist after accretion, as well as the role of magnetic fields in modifying vertical and horizontal diffusion, are both crucial for correctly interpreting observed atmospheric compositions and, by extension, the nature of the accreted planetary material. Addressing these issues will require improved modelling of magneto-convective diffusion and dedicated monitoring campaigns to map both magnetic topology and surface abundance evolution across a broader population of metal-enriched magnetic white dwarfs.

\begin{acknowledgements}
Based on observations collected at ESO's Paranal Observatory under programme IDs 115.27UN.002, 116.2921.001, and 116.2ASX.001 (PI: Bagnulo);  0102.D-0045(A), 0103.D-0029(A) and (B) (PI: Landstreet). All spectra are available in the observatory archives. JDL acknowledges the financial support of the Natural Sciences and Engineering Research Council of Canada, funding reference number 6377-2016. CPF received funding from the European Union's Horizon Europe research and innovation programme under grant agreement No. 101079231 (EXOHOST), and from the United Kingdom Research and Innovation Horizon Europe Guarantee Scheme (grant number 10051045). The Gravitational-wave Optical Transient Observer (GOTO) project
acknowledges the support of the Monash-Warwick Alliance; University of
Warwick; Monash University; University of Sheffield; University of
Leicester; Armagh Observatory \& Planetarium; the National Astronomical
Research Institute of Thailand (NARIT); Instituto de Astrofísica de
Canarias (IAC); University of Portsmouth; University of Turku;
University of Birmingham; and the UK Science and Technology Facilities
Council (STFC, grant numbers ST/T007184/1, ST/T003103/1 and
ST/Z000165/1). This work has made use of data from the Asteroid
Terrestrial-impact Last Alert System (ATLAS) project which is primarily
funded to search for near-Earth asteroids through NASA grants
NN12AR55G, 80NSSC18K0284, and 80NSSC18K1575; by-products of the NEO
search include images and catalogs from the survey area. This work was
partially funded by Kepler/K2 grant J1944/80NSSC19K0112 and HST
GO-15889, and STFC grants ST/T000198/1 and ST/S006109/1. The ATLAS
science products have been made possible through the contributions of
the University of Hawaii Institute for Astronomy, the Queen’s
University Belfast, the Space Telescope Science Institute, the South
African Astronomical Observatory, and The Millennium Institute of
Astrophysics (MAS), Chile. ZTF data were obtained with the Samuel
Oschin Telescope 48-inch and the 60-inch Telescope at the Palomar Observatory as part of the Zwicky Transient Facility project.
ZTF is supported by the National Science Foundation under Grants
No. AST-1440341 and AST-2034437 and a collaboration including
current partners Caltech, IPAC, the Weizmann Institute for Science,
the Oskar Klein Center at Stockholm University, the University of Maryland, Deutsches Elektronen-Synchrotron and Humboldt University, the TANGO Consortium of Taiwan, the University of Wisconsin at Milwaukee, Trinity College Dublin, Lawrence Livermore
National Laboratories, IN2P3, University of Warwick, Ruhr University Bochum, Northwestern University and former partners the
University of Washington, Los Alamos National Laboratories, and
Lawrence Berkeley National Laboratories. Operations are conducted
by COO, IPAC, and UW.
\end{acknowledgements}

\bibliography{sbabib}
\appendix

%\onecolumn
\section{Equivalent width measurements}\label{App_Log}
\begin{table*}
\caption{\label{Table_EW_WD1532}
WD\,1532$+$129: \bz\ values and EWs in various wavelength ranges from all spectra obtained with FORS2 and the 1200B grism.}
\centering
\begin{tabular}{cr@{$\pm$}lr@{$\pm$}lr@{$\pm$}lr@{$\pm$}l}
\hline\hline
 & \multicolumn{2}{c}{\bz} &
 \multicolumn{2}{c}{EW$_1$ (\AA)} &
 \multicolumn{2}{c}{EW$_2$ (\AA)} &
 \multicolumn{2}{c}{EW$_3$ (\AA)} \\
Date &
\multicolumn{2}{c}{(kG)} &
\multicolumn{2}{c}{4357--4421\,\AA} &
\multicolumn{2}{c}{4180--4335\,\AA} &
\multicolumn{2}{c}{4162--4466\,\AA} \\
\hline
2019-03-18 & $ -5.0$ & 1.4 & 16.541 & 0.039 & 56.898 & 0.069 & 77.850 & 0.096 \\
2019-06-19 & $-23.1$ & 2.1 & 17.664 & 0.069 & 58.350 & 0.107 & 82.182 & 0.148 \\
2025-04-03 & $+24.1$ & 1.0 & 19.175 & 0.029 & 64.191 & 0.050 & 90.050 & 0.072 \\
2025-04-05 & $+24.4$ & 0.9 & 19.279 & 0.029 & 65.054 & 0.048 & 90.990 & 0.069 \\
2025-04-23 & $+26.4$ & 1.2 & 19.375 & 0.034 & 64.069 & 0.058 & 89.734 & 0.082 \\
2025-04-27 & $+25.6$ & 0.9 & 19.165 & 0.029 & 64.064 & 0.050 & 89.879 & 0.071 \\
2026-04-18 & $ +1.6$ & 1.1 & 16.356 & 0.031 & 57.129 & 0.053 & 78.377 & 0.075 \\
\hline
\end{tabular}
\end{table*}

Table~\ref{Table_EW_WD1532} reports the EW measurements from FORS2 spectra.
The ISIS spectrum was not used for EW measurements because its EWs are affected by an instrumental bias that is not negligible compared with the variability detected in the FORS2 spectra.

%\newpage
\section{Can the observed line variability be due to scattered light?}\label{App_Scattered}
Since the observed variability of the line strength is quite subtle ($\sim 15\%$), it is important to seriously examine the possibility that it is of instrumental origin. 

All observations in which the EWs reach their maxima were obtained under dark conditions. By contrast, two observations showing weaker lines were taken in bright time. In particular, on March 18, 2019, when the line EWs were close to their minimum values, the observations were obtained with a fractional lunar illumination (FLI) of 90\%, with the Moon located $86^\circ$ from the target and at an elevation of approximately $18^\circ$. On June 19, 2019, when the line EWs were slightly larger but still below their maximum values, the observing conditions were even less favourable: the FLI was 97\%, with the Moon about $66^\circ$ from the target. One could suspect that scattered light was responsible for partially filling the spectral lines and making them appear weaker than when observed in dark time.

To rule out any contamination, we ensured that an observation taken at an epoch when a line-strength minimum and null \bz\ were predicted would be obtained during dark time. Data collected in 2026 confirm that the lines are weaker when the longitudinal magnetic field is zero, even in the absence of moonlight. In fact, intensity spectra obtained on 2026-04-18 (black solid line in Fig.~\ref{Fig_WD1532}) and on 2019-03-18 (black dotted line in Fig.~\ref{Fig_WD1532}), when $\vert \bz \vert $ was close to zero, are virtually indistinguishable. Similarly, all spectra obtained in April 2025, when \bz\ was near its positive maximum, are virtually indistinguishable from one another. These tests confirm that the instrument is stable, that data reduction was carried out in a fully consistent way, and that line variability is intrinsic to the source.

\begin{figure}
\begin{center}
\includegraphics[width=\columnwidth]{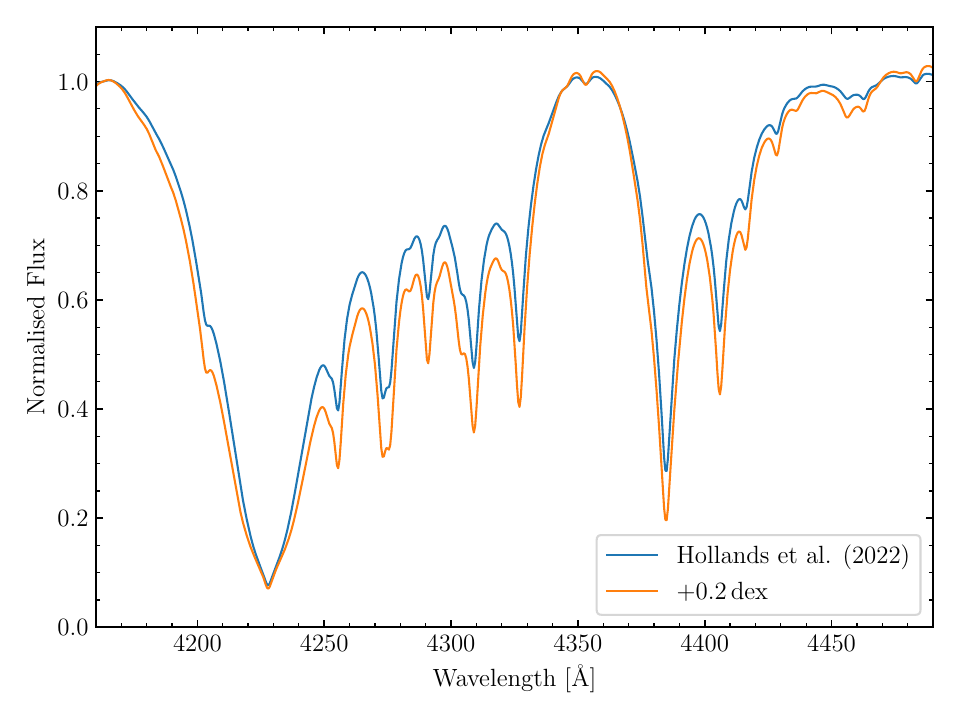}
\end{center}
\caption{\label{Fig_RT} Two theoretical spectra showing how an abundance change of +0.2 dex for both Fe and Ca produces a substantial line strength increase in the weaker Fe lines but onlt a small change in the stronger Ca\,{\sc i} line, as described in the text.
}
\end{figure}

The strong Ca\,{\sc i} 4227 line does not show a significant change in core depth, although Fig.~\ref{Fig_WD1532} hints at a slight strengthening or broadening of the wings when the weaker metal lines are strongest. This is naturally explained if the variability is due to abundance changes. Since the line core is nearly saturated, a modest increase in Ca abundance produces little extra absorption at line centre, whereas the unsaturated wings do respond and become slightly deeper, yielding only a modest increase in equivalent width. This behaviour is reproduced by the theoretical spectra shown in Fig.~\ref{Fig_RT}, both for the model based on \citet{Holetal22} (blue line) and for the test case in which the Fe and Ca abundances are increased by +0.2 dex (orange line): weaker lines strengthen markedly, while the saturated Ca\,{\sc i} 4227 line changes only weakly, mainly in its wings \citep[radiative transfer calculations were performed using the atmospheric models described by][]{Koester10}. This supports an abundance-based interpretation. In contrast, the amount of scattered light needed to explain the variability of the Fe lines would also be expected to alter the Ca\,{\sc i} 4227 profile much more noticeably, which is not seen.
%%%%%%%%%%%%%%%%%%%%%%%%%%%%%%%%%%%%%%%%%%%%%%%%%%%%%%%%%%%%%%%%%%%%%%%%%%%%%%%%%%%%%%%%%%%%%
%%%%%%%%%%%%%%%%%%%%%%%%%%%%%%%%%%%%%%%%%%%%%%%%%%%%%%%%%%%%%%%%%%%%%%%%%%%%%%%%%%%%%%%%%%%%%
\newpage
\section{Period analysis}\label{App_Period}
\begin{figure}[h!]
\begin{center}
\includegraphics[width=\columnwidth]{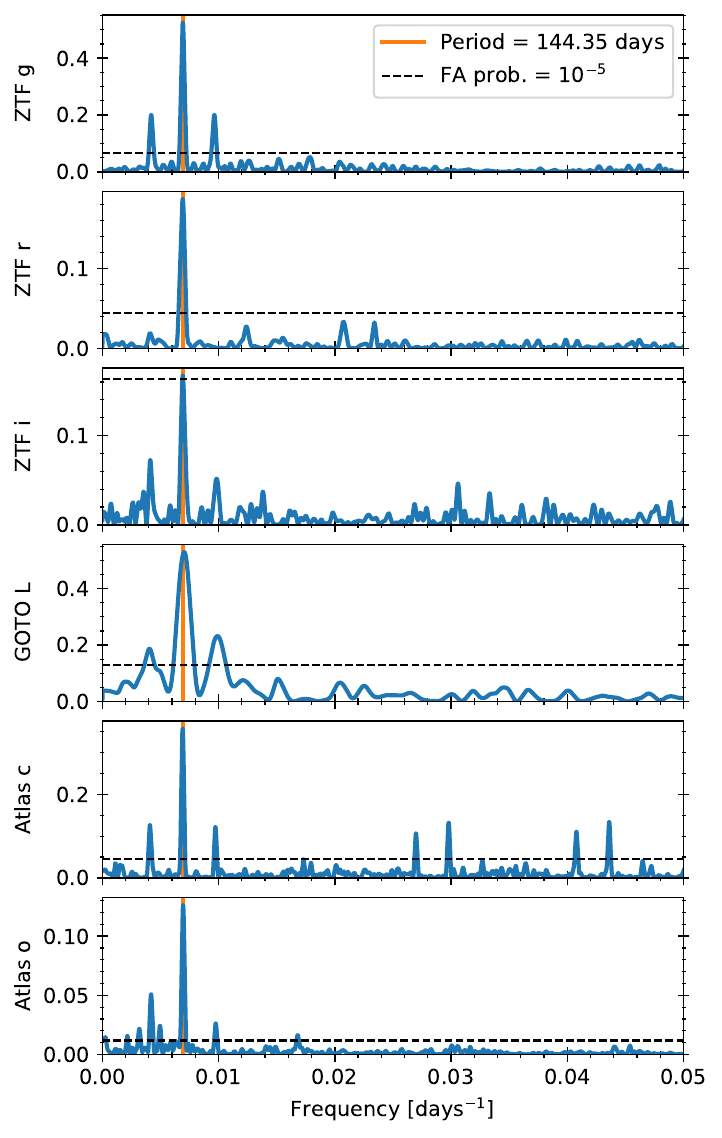}
\end{center}
\caption{\label{Fig_WD1532_power} Lomb-Scargle periodograms of the photometric data of WD\,1532+129. The vertical orange line marks the frequency corresponding to a period of 144.35 days. The dashed horizontal line shows the height corresponding to a false alarm probability of $10^{-5}$. 
}
\end{figure}

\begin{figure}[h]
\begin{center}
\includegraphics[width=7.5cm,trim={0.25cm 0.20cm 0.25cm 0.1cm},clip]{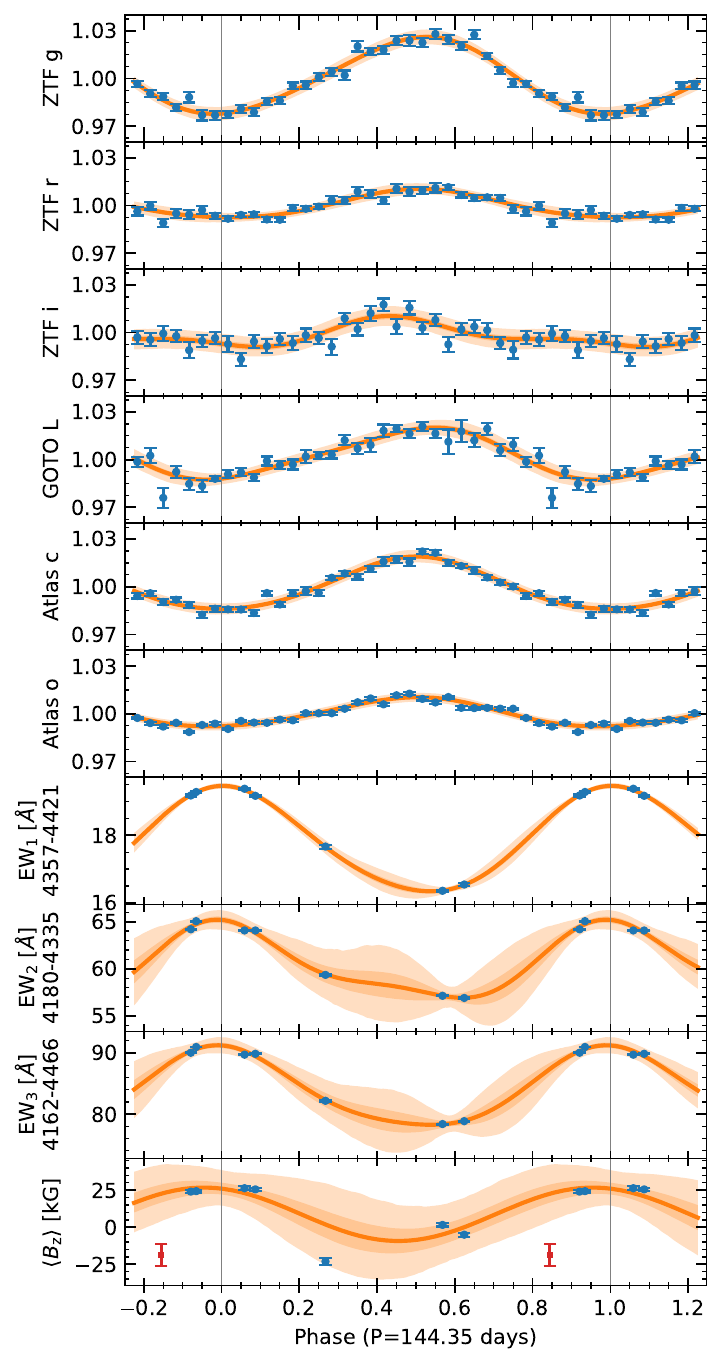}
\end{center}
\caption{\label{Fig_WD1532_all144d} Observed quantities for WD\,1532$+$129 phased with the best photometric period of 144.35\,d and a zero-point of MJD 59047.566. Orange curves show the corresponding best-fitting models, and the shaded regions show the 1\,$\sigma$ and 3\,$\sigma$ confidence intervals. This period is inconsistent with the observed mean longitudinal field variations shown in the bottom panel. The red point with visible error bars in the bottom panel represents the \bz\ estimate obtained with ISIS at the WHT, which was not included in the fit. The vertical lines delimit one complete rotational cycle.
}
\end{figure}

Figure~\ref{Fig_WD1532_power} shows the Lomb-Scargle periodograms of the photometric data of WD\,1532+129. Figure~\ref{Fig_WD1532_all144d} shows the best fits to all quantities obtained after adopting the dominant period from the photometric data ($P=144.35$\,d). It is clear that this value of the rotation period does not fit the mean longitudinal field data. 

%%%%%%%%%%%%%%%%%%%%%%%%%%%%%%%%%%%%%%%%%%%%%%%%%%%%%%%%%%%%%%%%%%%%%%%%%%%%%%%%%%%%%%%%%%%%%
%%%%%%%%%%%%%%%%%%%%%%%%%%%%%%%%%%%%%%%%%%%%%%%%%%%%%%%%%%%%%%%%%%%%%%%%%%%%%%%%%%%%%%%%%%%%%

\end{document}